\documentclass[aps,prd,reprint,superscriptaddress,longbibliography,nofootinbib,floatfix,showpacs]{revtex4-1}

\usepackage{epsfig}
\usepackage{amsmath,amssymb,amsfonts}
\usepackage{hyperref}
\usepackage{mathrsfs}
\usepackage{slashed}
\usepackage{graphicx}
\usepackage{verbatim}
\usepackage[usenames]{color}

\newcommand{\be}{\begin{equation}}
\newcommand{\ee}{\end{equation}}
\newcommand{\bi}{\begin{itemize}}
\newcommand{\ei}{\end{itemize}}
\newcommand{\bea}{\begin{eqnarray}}
\newcommand{\eea}{\end{eqnarray}}

\newcommand{\bra}[1]{\langle\,#1\,|}          
\newcommand{\ket}[1]{|\,#1\,\rangle}          

\newcommand{\ud}{\mathrm{d}}

\newcommand{\LCm}{{\scriptscriptstyle -}} 
\newcommand{\LCp}{{\scriptscriptstyle +}}
\newcommand{\LCpm}{{\scriptscriptstyle \pm}}
\newcommand{\LCmp}{{\scriptscriptstyle \mp}}

\newcommand{\LCperp}{{\scriptscriptstyle \perp}}

\begin{document}

\title{Pair production: the view from the lightfront}

\author{Florian Hebenstreit}
\email[]{florian.hebenstreit@uni-graz.at}
\affiliation{Department of Physics, Ume\aa\ University, SE-901 87 Ume\aa, Sweden}
\affiliation{Institut f\"ur Physik, Karl-Franzens Universit\"at Graz, A-8010 Graz, Austria}

\author{Anton Ilderton}
\email[]{anton.ilderton@physics.umu.se}
\affiliation{Department of Physics, Ume\aa\ University, SE-901 87 Ume\aa, Sweden}

\author{Mattias Marklund}
\email[]{mattias.marklund@physics.umu.se}
\affiliation{Department of Physics, Ume\aa\ University, SE-901 87 Ume\aa, Sweden}

\begin{abstract}
We give an exact, analytic, and manifestly gauge invariant account of pair production in combined longitudinal and transverse electromagnetic fields, both depending arbitrarily on lightfront time. The instantaneous, nonperturbative probability of pair creation is given explicitly along with the spectra of the final particle yield. Our results are relevant to high-intensity QED experiments now being planned for future optical and x-ray free electron lasers.

\end{abstract}
\pacs{}
\maketitle

\section{Introduction}
The use of laser light sources in examining the high intensity regime of QED continues to draw attention, prompted by the advent of a new generation of both x-ray and optical laser facilities such as the European XFEL and the Extreme Light Infrastructure project~\cite{XFEL,ELI}. Dominating the theoretical activity in this area is the pursuit of (Schwinger) pair production in strong background fields \cite{Dunne:2010zz}.  

In this paper we apply the Dirac-Heisenberg-Wigner (DHW) formalism \cite{original, orig2,orig3} to the problem of pair production in background fields. Within this approach, which was developed in the early 1990's \cite{BialynickiBirula:1991tx,Zhuang:1995pd} and has gained increased attention in recent years \cite{Hebenstreit:2010vz,Hebenstreit:2010cc, BialynickiBirula:2011uh, Hebenstreit:2011wk}, one studies phase space distribution functions instead of the usual S-matrix elements. Since the DHW formalism deals essentially with quasi-probabilities, interpretation can be challenging. Nevertheless, quite some progress can be made by considering simple, but nontrivial, cases.

An advantage of this approach is that the key object, called the DHW function, which is essentially the field density in an appropriate state, can in principle be obtained by solving a single partial differential equation. This is true at least when the gauge field is external: going beyond this is notoriously hard within the DHW approach \cite{Blaschke:2011is}, while perturbation theory is obviously straightforward in field theory. In order to ensure that the obtained solution is physical, one must use sensible initial data which corresponds to calculating the DHW function in a physical state. Alternatively, one could try to construct the DHW function directly. This is challenging since it requires finding the exact solutions of the Dirac equation in the chosen background field and then quantising the theory.

In this paper we continue the investigation of the DHW function started in \cite{Hebenstreit:2010cc} using lightfront methods (see \cite{Heinzl:2010vg,Honkanen:2010rc, Meuren:2011hv} for applications of related methods to QED in a variety of strong external fields). Our previous results focussed on plane wave backgrounds, i.e.\ transverse, orthogonal electric and magnetic fields of equal magnitude, depending on lightfront time $x^\LCp$. While we made progress in understanding the effective mass of a particle in an arbitrary pulse, we were of course unable to study pair production since single plane waves cannot produce pairs. In this paper we retain the plane wave fields but add a longitudinal electric field. Both of our fields will depend arbitrarily on lightfront time, allowing us to model modern short-duration laser pulses \cite{Heinzl:2010vg,Heinzl:2009nd,Mackenroth:2010jk,Mackenroth:2010jr}.

To obtain the DHW function, we follow the second approach described above: we will therefore present new solutions of the Dirac equation in a combination of longitudinal and transverse fields, quantise the theory and calculate the lightfront DHW function directly. This function, as we will show, can be interpreted as a (quasi)probabilistic measure of electron/positron occupation numbers. This will give us a clear signal of pair creation as we will be able to see, in a gauge invariant manner, the filling of states as particles are produced. We also confirm the results given in \cite{Tomaras:2000ag,Tomaras:2001vs,Woodard:2001ai,Woodard:2001hi}, namely that from the infinitely boosted lightfront frame one sees only the created positrons (modulo the choice of field) since the electrons decouple from the theory after creation.

We begin in Sect.~\ref{WigSect} by reviewing the DHW approach and presenting some basic results. In Sect.~\ref{TechSect} we give the required solution of the Dirac equation in our chosen background. The quantisation of the theory and construction of the DHW function is not too hard but the expressions involved can become quite lengthy, and are therefore relegated to the appendices. We give the exact DHW function in Sect.~\ref{PairSect} and analyse pair creation on the lightfront, using explicit examples of both short and long pulses. We also reconstruct the final particle spectrum in the lab frame. Conclusions are given in Sect.~\ref{ConcSect}. Our lightfront conventions are quite standard but we encourage the reader not familiar with lightfront methods to consult the appendices for details.

\section{The lightfront DHW function} \label{WigSect}
The DHW function is essentially given by the Fourier transform of the fermion field density in a chosen state, usually the vacuum. Our background fields will depend on lightfront time $x^\LCp\equiv (x^0+x^3)/\sqrt{2}$ and so it is natural to work in lightfront field theory. On the lightfront, the fermion fields $\psi$ split into dynamical fields $\psi_\LCp$, and constrained fields $\psi_\LCm$ (this is reviewed below, details are not needed here). It is therefore convenient, and simpler, to study the DHW function defined by the density of the dynamical fields rather than the full Dirac spinor.

The equal lightfront time DHW function begins with the dynamical fermion density $U$ in, say, the vacuum $\ket{0}$. Noting that we use a sans-serif font to denote the spatial lightfront variables and momenta, i.e.\ ${\sf x}\equiv \{x^\LCm,x^\LCperp\}$ and ${\sf p}\equiv \{p_\LCm,p_\LCperp\}$, this density is
\be\label{Udef}
	U_{\alpha\beta}\equiv \bra{0}\big[\psi_{\LCp\alpha}(x^\LCp,\mathsf{x}_2), \psi^\dagger_{\LCp\beta}(x^\LCp,{\sf x}_1)\big]\ket{0} \;,
\ee
where $\alpha$ and $\beta$ are spin indices.  Setting ${\sf x}_2 \equiv {\sf x}+{\sf y}/2$ and ${\sf x}_1 \equiv {\sf x}-{\sf y}/2$, the DHW function is defined by Fourier transforming with respect to the relative co-ordinate $\sf y$:
\be \label{defn}\begin{split}
	W^\LCp_{\alpha\beta}(x^\LCp; &{\sf x},\mathsf{p}) = \sqrt{2}\!\int\!\ud^3 \mathsf{y}\ e^{i\mathsf{p}.\mathsf{y} + ie \int \ud z.A(z)} U_{\alpha\beta}\;,
\end{split}
\ee
where gauge invariance of $W^\LCp$ is ensured by the Wilson line in the exponent. The line integral is taken over the straight path from $\sf{x}_1$ to $\sf{x}_2$ which corresponds to minimally coupling the free DHW function by replacing $\partial\to D$ \cite{Elze:1986qd}. The factor of $\sqrt{2}$ is a normalisation. Note that, being gauge invariant, the DHW function can be calculated in any gauge. Some intuition for what the DHW function represents can be built up by calculating it in a variety of simple states, to which we now turn.

\subsection{Free theory}
\begin{figure}[t!]
\centering \includegraphics[width=0.8\columnwidth]{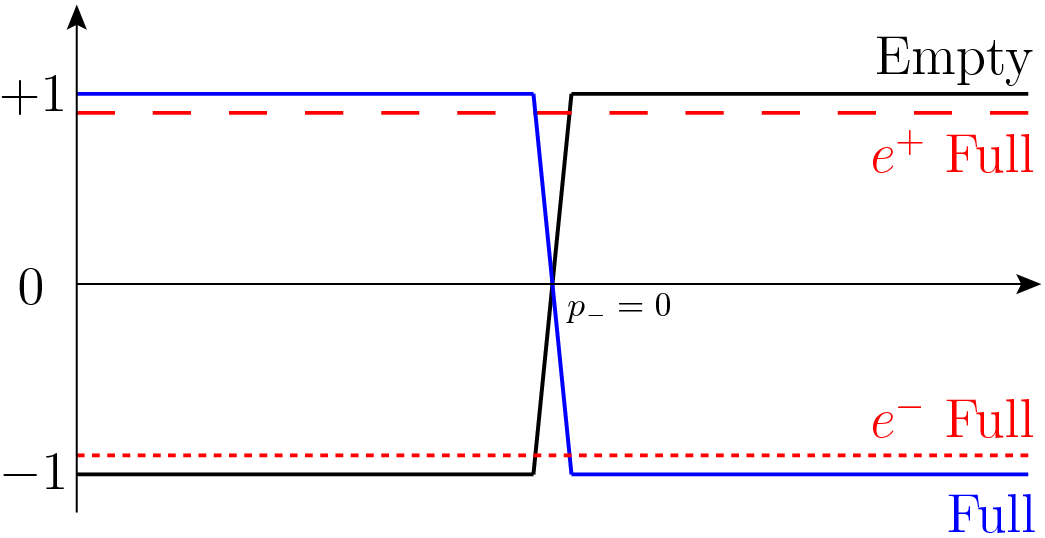}
\caption{\label{Snygga} The free $W$ as a function of $p_\LCm$, calculated in the ordinary, empty vacuum (black), in the state filled with positrons (red, dashed), the state filled with electrons (red, dotted) and the completely filled state (blue). The region $p_\LCm<0$ to the left of the plot is controlled by positrons, $p_\LCm>0$ by electrons.}
\end{figure}

\begin{figure}[t!]
\centering \includegraphics[width=0.8\columnwidth]{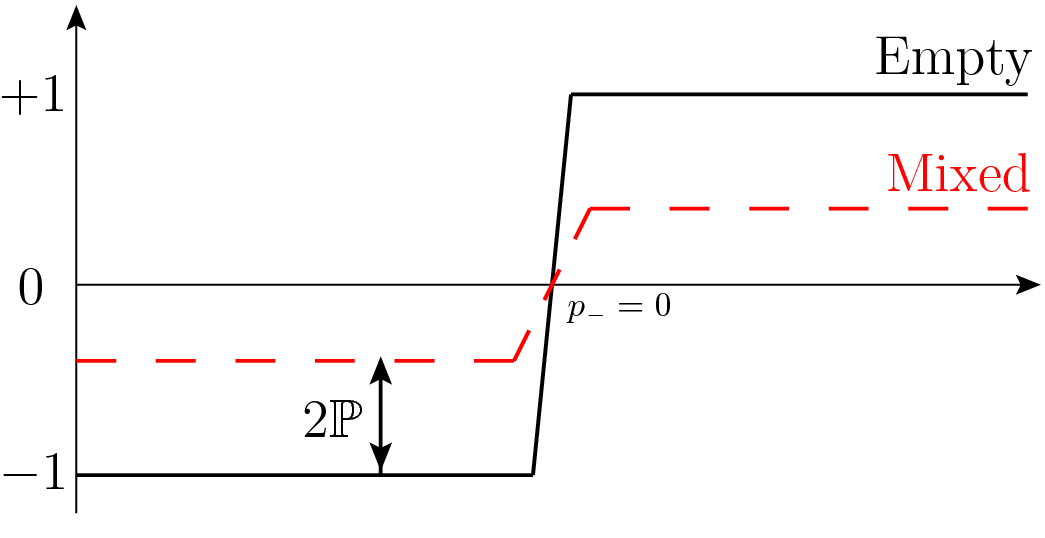}
\caption{\label{Snygga2} The DHW function in the mixed state, (\ref{MixedWigner}). The deviation from the vacuum DHW function is given by twice the probability for being in the state $\ket{\text{full}}$.}
\end{figure}

All our DHW functions will be proportional to the lightfront projector $\Lambda^\LCp\equiv \tfrac{1}{2}\gamma^\LCm\gamma^\LCp$, so we write
\be
	W^\LCp_{\alpha\beta} \equiv \Lambda^\LCp_{\alpha\beta} W\;.
\ee
The DHW function (\ref{defn}) for free fermions is easily found by writing down the mode expansion, calculating the expectation value and performing the Wigner transformation (without Wilson line). One finds
\be\label{Sign}
	W(x^\LCp; \mathsf{x},\mathsf{p}) = \text{Sign}(p_\LCm) \;,
\ee
which is spatially homogeneous and displays only a simple dependence on the lightfront momentum $p_\LCm$. This behaviour is due to the existence of both positrons and electrons. To see why, replace $\ket{0}$ in (\ref{Udef}) with $\ket{\text{full}}$, in which every positron and electron state is occupied. The DHW function becomes
\be\label{FilledWigner}
	W(x^\LCp; \mathsf{x},\mathsf{p}) = \text{Sign}(-p_\LCm) \;, \quad \text{filled vacuum}\;.
\ee
Similarly, filling all the electron {\it or} positron states, one obtains instead
\be\begin{split}\label{EitherFilled}
	W(x^\LCp; \mathsf{x},\mathsf{p}) &= -1 \;, \quad \text{electrons filled}\;, \\
	&= +1 \;, \quad \text{positrons filled} \;.
\end{split}\ee
These results are shown in Fig.~\ref{Snygga}. The region $p_\LCm>0$ is controlled by electrons and $p_\LCm<0$ by positrons: this is a matter of convention (it does not refer to negative energy) which follows from the choice of exponent in the transform (\ref{defn}), since the mode expansion for $\psi$ looks like
\be
	\psi \sim \int\limits_0^\infty\frac{\ud k_\LCm}{k_\LCm} \int \ud^2 k_\LCperp\ (e^{-ik.x}u^s_k b^s_k + e^{ik.x} v^s_k d_k^{s\dagger}).
\ee
Following this,  the DHW variable $\sf p$ may be associated with the momentum of an electron, or {\it minus} the momentum of a positron. Consider also a mixed state,
\be\label{MixedState}
	\ket{\text{mixed}} =\sqrt{1-\mathbb{P}}\ket{0} +  \sqrt{\mathbb{P}}\ket{\text{full}} \;,
\ee
which has probability $\mathbb{P}$ of being full and $1-\mathbb{P}$ of being empty. The DHW function is easily found to be
\be\label{MixedWigner}\begin{split}
	W &= \text{Sign}(p_\LCm)+2\mathbb{P}\,\text{Sign}(-p_\LCm) \;,
\end{split}\ee
so that, for example, the $p_\LCm<0$ portion of the DHW function is raised by a height $2\mathbb{P}$ relative to that in the vacuum, see Fig.~\ref{Snygga2}. The DHW function therefore gives us a (quasi)probabilistic measure of the occupation numbers of electrons and positrons. The above points are worth keeping in mind for later, as they will aid our interpretation of the DHW function in the interacting theory. 

\subsection{Plane wave backgrounds}
The DHW function for both scalars and spinors in an arbitrary plane wave background was calculated in \cite{Hebenstreit:2010cc}. That paper also considered the covariant DHW function in which the lightfront times are also separated, i.e.\ one works with the density $\big[\psi_\LCp(x_2),\psi^\dagger_\LCp(x_1)\big]$ with ${x^\mu_{1,2} = x^\mu \pm y^\mu/2}$. The DHW function is then defined by an integral over $\ud^4 y$, and was found to be
\be
	\text{Sign}(p_\LCm)\int\!\ud y^\LCp \exp\bigg[i y^\LCp\bigg(p_\LCp - \frac{p_\LCperp^2+ M^2(x^\LCp,y^\LCp)}{2p_\LCm}\bigg)\bigg] \;,
\ee
where $M^2$ is clearly an effective mass (see \cite{Hebenstreit:2010cc,Kibble:1975vz} for details) which extends the intensity-dependent mass shift from purely periodic plane waves \cite{Nikishov:1964zza,Narozhnyi:1964,Narozhnyi:1976zs}, to arbitrary plane wave backgrounds. Integrating out $p_\LCp$, one recovers, by definition, our $W$: this is precisely the same as in the free theory (\ref{Sign}). The natural interpretation of this result is the well known statement that plane waves do not create pairs.

In the following sections we will examine what happens to the DHW function when a pair-creating longitudinal electric field is added to the plane waves. We will continue to focus on the simpler $W$ rather than the covariant DHW function as this is the more common approach in the literature, and because, following the above, any deviation in $W$ from the free vacuum result (\ref{Sign}) must be due (at least in part) to the longitudinal field.

\section{Longitudinal and transverse fields}\label{TechSect}
We consider a laser pulse moving up the $x^3$-axis. This defines our `longitudinal' direction, while ${x^1,x^2}$ are the transverse directions. A variety of models for the laser field can be found in the literature: the case of a constant, longitudinal electric field is of course covered by Schwinger's classic result \cite{Schwinger:1951nm}. The models for which most analytic progress can be made (in terms of calculating scattering amplitudes) are plane waves depending on $x^\LCp$ \cite{Nikishov:1964zza,Narozhnyi:1964,Narozhnyi:1976zs}. The combination of a constant longitudinal electric field and periodic plane waves was described in \cite{Narozhnyi:1976zs}. The case of a purely longitudinal electric field $E(x^\LCp)$ depending arbitrarily on $x^\LCp$ was covered by \cite{Tomaras:2000ag,Tomaras:2001vs} (and includes Schwinger's result as a particular case). The lightfront methods used in those papers can be extended to cover the case of longitudinal $E(x^\LCp)$ with a parallel B-field also depending on $x^\LCp$ \cite{Soussa:2002ed}. For fields depending on both $x^\LCp$ and $x^\LCm$ see \cite{Fried:2001ga,Avan:2002dn}. Purely longitudinal electric fields $E(x^0)$ depending arbitrarily on (instant) time $x^0$ are widely used in the literature. Such fields model the focus of counter propagating laser pulses in which the magnetic field components cancel. They have been used to investigate pair production for oscillating fields \cite{Brezin:1970xf,Alkofer:2001ik,Blaschke:2005hs}, pulsed fields \cite{Narozhnyi:1970uv,Popov:1972} and pulsed fields with sub-cycle structure~\cite{Popov:2001ak, Hebenstreit:2009km, Dumlu:2010ua}.

Here we further extend the above results, covering the case of longitudinal electric and transverse electromagnetic plane wave fields, both depending arbitrarily on~$x^\LCp$. We work in `anti-lightcone' gauge $A^\LCm \equiv A_\LCp = 0$ (the usual lightcone gauge is $A_\LCm=0$). The remaining components of the potential are given by,
\be\label{TheFields}
	A_\LCm = -\int\limits_0^{x^\LCp}\!\ud y\ E^{\scriptscriptstyle\parallel}(y) \;,\qquad A_\LCperp = \sqrt{2}\int\limits_0^{x^\LCp}\!\ud y\ E^\LCperp(y) \;,
\ee
where we assume for simplicity that the fields turn on at $x^\LCp=0$. This can, and will, be relaxed below.

\subsection{Solutions of the Dirac equation}
Defining the projectors $\Lambda^{\LCpm}\equiv\tfrac{1}{2}\gamma^{\LCmp}\gamma^{\LCpm}$ the fermion field decomposes into $\psi \equiv \psi_\LCp + \psi_\LCm$ with $\psi_\LCpm \equiv \Lambda^\LCpm\psi$. The Dirac equation then separates into
\bea
	\label{Dirac1} i\partial_\LCp\psi_\LCp = \tfrac{1}{2}(i\gamma^\perp D_\LCperp + m)\gamma^\LCm \psi_\LCm \;, \\
	\label{Dirac2} iD_\LCm\psi_\LCm = \tfrac{1}{2}(i\gamma^\perp D_\LCperp + m)\gamma^\LCp \psi_\LCp \;.
\eea
We immediately take the Fourier transform of the {\it transverse} coordinates, $i\partial_\LCperp \to k_\LCperp$, which replaces
\be
	iD_\LCperp \to k_\LCperp - e A_\LCperp(x^\LCp) \equiv \pi_\LCperp(x^\LCp) \;.
\ee
One solves the Dirac equation by first observing that $\psi_\LCm$ is a constrained field, since it can be expressed in terms of $\psi_\LCp$ using (\ref{Dirac1}):
\be\label{MinusFromPlus}
	\psi_\LCm \equiv \frac{\gamma^\LCperp\pi_\LCperp+m}{\omega^2}\ i\gamma^\LCp\partial_\LCp \psi_\LCp \;,
\ee
where the mode frequency is defined by
\be\begin{split}
	\omega^2(x^\LCperp) &\equiv \pi^2_\LCperp(x^\LCp)+m^2 \\
	&= k_\LCperp^2 + m^2 + e^2A_\LCperp^2(x^\LCp) - 2e k_\LCperp A_\LCperp(x^\LCp) \;,
\end{split}\ee
in which we recognise the Volkov exponent \cite{Volkov:1935}. 
Substituting (\ref{MinusFromPlus}) into (\ref{Dirac2}), and noting that $D_\LCm$ and $\partial_\LCp$ do not commute, one obtains a simple equation for $\psi_\LCp$:
\be\label{SimpleDirac}
	 D_\LCm \partial_\LCp \psi_\LCp = -\frac{\omega^2}{2}\psi_\LCp\;.
\ee
If we try to Fourier transform $i\partial_\LCm\to k_\LCm$, we see that solving (\ref{SimpleDirac}) requires inverting
\be\label{HelloZero}
	k_\LCm - eA_\LCm(x^\LCp) \;,
\ee
which can clearly vanish, possibly multiple times, for a given $k_\LCm$. This is the zero-mode problem of lightfront field theory  \cite{Heinzl:2000ht}, but made time-dependent by the external field. The physics of the zero-mode in the current context is as follows. An electron with momentum $k_\LCm$ at time $x^\LCp=0$ acquires (as follows from solving the Lorentz equation) a momentum $k'_\LCm = k_\LCm - eA_\LCm(x^\LCp)$ at later times. We have $e<0$, so if we imagine that $E^{\scriptscriptstyle\parallel}$ is positive, then $k'_\LCm$ will vanish at some later lightfront time, which means the electron moves parallel to the $x^\LCm$ axis (reaches the speed of light) at this instant: it therefore vanishes from the theory since it cannot be seen at any subsequent lightfront time \cite{Woodard:2001ai,Woodard:2001hi}. Note that a positron's momentum, on the other hand, only increases in the above circumstances. Hence the positrons remain in the theory. This will be useful for later. 

\begin{figure}[t!]
\includegraphics[width=0.75\columnwidth]{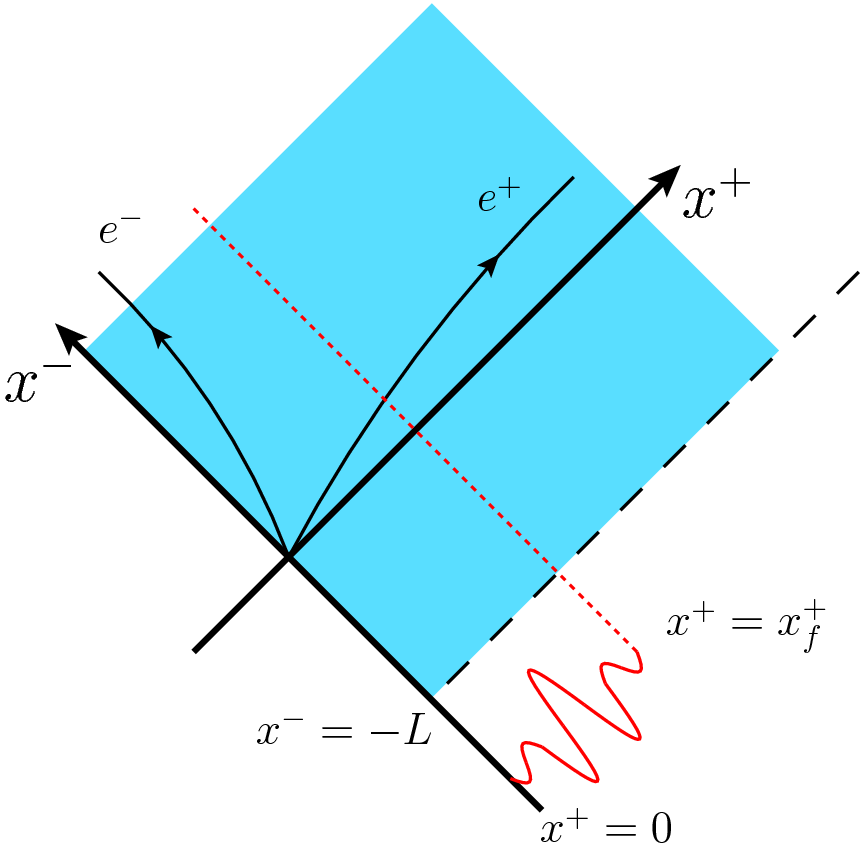}
\caption{\label{FIG1} The domain of our solution (\ref{this}), $x^\LCp>0$ and $x^\LCm>-L$. The finite-duration background fields depend on $x^\LCp\in 0\ldots x^\LCp_f$, as is also illustrated, along with the behaviour of electrons and positrons created within the field.}
\end{figure}
We are now ready to give the solution of the Dirac equation. We follow the method of \cite{Tomaras:2000ag,Tomaras:2001vs}. The idea is to turn the fields on at $x^\LCp=0$ (for convenience, this can be relaxed, see below), and solve the Dirac equation in the semi-infinite region $x^\LCp>0$, $x^\LCm>-L$ for some positive $L$, in terms of initial data. This gives, as we will see, a prescription for handling the singularity at $k_\LCm-eA_\LCm=0$. In the end the limit $L\to\infty$ is taken.  The domain of our solution, together with an illustration of our fields and the motion of particles within them, is shown in Fig.~\ref{FIG1}.

It may be checked directly that the solution to (\ref{Dirac1})-(\ref{Dirac2}) in $x^\LCp>0$, $x^\LCm>-L$ is
\bea\nonumber
	&\psi_\LCp(x^\LCp,x^\LCm) = \displaystyle\int\limits_{-L}^\infty\!\ud y^\LCm\ \psi_\LCp(0,y^\LCm) \bar{D}_\LCm(y)G\big( {x^\LCp},{0};{x^\LCm},{y^\LCm}\big) \\
	\label{this} &-\displaystyle \int\limits_{0}^{x^\LCp}\!\ud y^\LCp\ \frac{\partial\psi_\LCp}{\partial y^\LCp}(y^\LCp,{ -L}) G\big({x^\LCp},{y^\LCp};{x^\LCm},{-L}\big)\;,
\eea
with $\psi_\LCm$ given by (\ref{MinusFromPlus}). We consider the various terms. First, $\psi_\LCp$'s dependence on the boundary data is explicit:  the solution depends on $\psi_\LCp$ on the characteristic $x^\LCp=0$ and $\partial_\LCp \psi _\LCp \sim \psi_\LCm$ on the characteristic $x^\LCm=-L$, since
\be
	i\partial_\LCp\psi_\LCp = \frac{1}{2}(\slashed{\pi}+m)\gamma^\LCm\psi_\LCm \;,
\ee
from (\ref{MinusFromPlus}). The function $G$ is
\be\begin{split}\label{G-DEF}
	G\big(&x^\LCp, y^\LCp; x^\LCm, y^\LCm \big) = \\
	&-i\,\int\!\frac{\ud k_\LCm}{2\pi}\ \frac{e^{i(y^\LCm-x^\LCm)(k_\LCm+i/L)}}{k_\LCm- eA_\LCm(y^\LCp)+i/L}\mathcal{E}_{k_\LCm}(y^\LCp,x^\LCp) \;,
\end{split}
\ee
and $\mathcal{E}$ is defined by
\be\label{E}
	\mathcal{E}_{k_\LCm}(x^\LCp,y^\LCp) = \exp\bigg[-\frac{i}{2}\int\limits_{y^\LCp}^{x^\LCp}\!\frac{\ud s\ \omega^2(s)}{k_\LCm-eA_\LCm(s)+i/L}\bigg] \;.
\ee
It is worth considering this function in a little detail, as it exhibits the essential difference between the transverse and longitudinal fields. The transverse plane wave fields enter just as in the Volkov solution, in the numerator of the exponent \cite{Volkov:1935}. These terms may therefore be recovered by resumming all orders of perturbation theory in the plane wave coupling $eA_\LCperp$. The longitudinal field, on the other hand, appears in the denominator and exhibits a singularity on the real line, regulated by the factors of $i/L$: when $L\to\infty$ this leads to an essential singularity in the coupling, as  in Schwinger's result.

An advantage of the approach we adopt is that the differences between the types of field, and the important structures, are laid bare. Nothing is hidden inside the behaviour of special functions, as is frequently the case in the instant-form approach: the equal $x^0$ (instant time)  DHW function is expressed in terms of parabolic cylinder functions for $E(x^0)=E$, constant, and in terms of hypergeometric functions for $E(x^0)=E\operatorname{sech}^2(\omega x^0)$ \cite{Hebenstreit:2010vz}. However, a disadvantage of our approach is that expressions quickly become lengthy. For this reason, the quantisation of (\ref{this}) and the calculation of the DHW function are left to the appendix. Related calculations are explicitly performed in \cite{Tomaras:2001vs}, which the reader may consult for further examples.  The final result for the DHW function in the limit $L\to\infty$ is, however, extremely compact, and we turn to it now.

\section{Pair production} \label{PairSect}
If one considers only a longitudinal electric field depending on $x^\LCp$, one finds that not only the vacuum persistence amplitude but also the pair production rate may be calculated instantaneously as a function of $x^\LCp$. The derivation of this latter result requires a careful interpretation of the Heisenberg operators in order to identify the pair production probability \cite{Tomaras:2000ag,Tomaras:2001vs}. We can provide a (positive) check of that interpretation using the DHW function. Moreover, our approach is manifestly gauge invariant. 

It was found in \cite{Narozhnyi:1976zs} that the addition of a plane wave to a {\it constant} electric field (Schwinger's case) does not change the vacuum persistence amplitude. We will see for our fields that this remains true: the plane wave has no impact on the {\it creation} of particles, nor the properties they are created with. Rather neatly, though, our results make clear the {\it post-creation} effect of the plane waves on the particles.

\subsection{The DHW function} \label{WignerSect}
Our solution of the Dirac equation (and its quantisation, as described in the appendix) is valid for arbitrarily longitudinal and plane wave fields. We present here results for the case in which $E^{\scriptscriptstyle\parallel}$ is assumed to be positive, so that $eA_\LCm$ is positive and increasing, as this is when the singularities in (\ref{E}) have the simplest structure in phase space.  Our longitudinal fields therefore model subcycle pulses, which are of considerable current interest \cite{Tajima}. The plane wave fields remain arbitrary.
   
We can now give the DHW function $W$. At $x^\LCp=0$, $W$ is that of the free theory, see (\ref{Sign}). Once the longitudinal fields turn off at $x^\LCp=x^\LCp_f$, the theory becomes stable against pair production and we find that the DHW function again becomes constant in lightfront time, matching its final value in the pulse. For the duration of the pulse, i.e. $0<x^\LCp<x^\LCp_f$, the DHW function is
\be\begin{split}\label{result}
	W(x^\LCp; &\mathsf{x},\mathsf{p}) = \text{Sign}(p_\LCm) + 2\,\mathbb{P}\,\theta(-p_\LCm)\theta\big(eA_\LCm(x^\LCp)+p_\LCm\big)\;. 
\end{split}
\ee
The first term is the DHW function of the empty vacuum. The second term contains the effects of the background fields, and has a form similar to that in (\ref{MixedWigner}), though it is restricted to $p_\LCm<0$. The description and investigation of this term, in particular $\mathbb{P}$, will occupy the remainder of the paper.

It is important before embarking on this to give the interpretation of the DHW variable $\sf p$. In the free theory, see Sect.~\ref{WigSect}, $\sf p$ ($-\sf p$) is the kinetic momentum of an electron (positron). This also holds once the background fields turn off and the theory again becomes free (the final particle spectrum is of course what we would be interested in experimentally). The DHW function smoothly connects the initial and final distributions in a gauge invariant manner. Furthermore, (\ref{defn}) shows that (canonical!) momentum dependence on, say, $\sf k$ in the density will be set equal to ${\sf p}+e{\sf A}$ in the Wigner function, so that $\sf p$ is naturally interpreted as a kinematic momentum. From here on we therefore associate $p_\LCm<0$ with a positron momentum $\pi_\LCm \equiv -p_\LCm>0$, as in the free theory: from (\ref{result}), this is clearly the region of interest.

\subsection{From dynamics to probabilities}
\begin{figure}[b!]
\includegraphics[width=0.8\columnwidth]{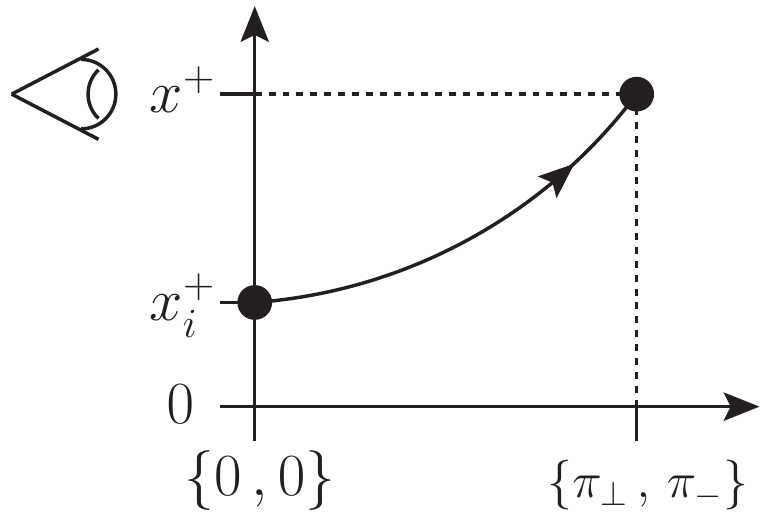}
\caption{\label{ParBildFig} A particle is created at time $x_i^\LCp$, with probability determined by the electric field strength at that time. The particle has zero longitudinal momentum, and transverse momentum normally distributed around zero. It is observed at a later time $x^\LCp$, after which it has acquired longitudinal and transverse momenta $\pi_\LCm$ and $\pi_\LCperp$. }
\end{figure}

In order to give the most compact and intutive expression for $\mathbb{P}$ it is useful to recall some results on the motion of particles in our background fields. The Lorentz equation for a positron with kinematic momentum $\pi_\mu$ and charge $-e>0$ is $\ud\pi_\mu = -e F_{\mu\nu}\ud x^\nu$. Suppose then, that  a positron is created with momentum $\pi_\LCm=0$ at some initial time $x_i^\LCp$. From the Lorentz equation, it will at a later time $x^\LCp$ have momentum
\be \label{HiddenInt0}
	\pi_\LCm = eA_\LCm(x^\LCp)-eA_\LCm(x^\LCp_i) \;,
\ee
using (\ref{TheFields}). Since $eA_\LCm$ is positive and increasing for the duration of the pulse, it has a unique inverse $X_p$ such that  $eA_\LCm(X_p) = p$ and $X_{eA_\LCm(x)}=x$. It follows that, on observation of a positron with momentum $\pi_\LCm$ at time $x^\LCp$, the `initial time' could be reconstructed from (\ref{HiddenInt0}):
\be\begin{split}\label{HiddenInt1}
	eA_\LCm(x^\LCp_i) &= eA_\LCm(x^\LCp)-\pi_\LCm \\
	\implies x^\LCp_i(\pi_\LCm) &\equiv X_{eA_\LCm(x^\LCp)-\pi_\LCm} \;.
\end{split}
\ee
(We suppress the dependence of $x^\LCp_i$ on $x^\LCp$ for compactness.) If the positron also has zero transverse momentum at the initial time, its later transverse momentum is
\be \label{HiddenInt2}
	\pi_\perp(\pi_\LCm) \equiv eA_\LCperp(x^\LCp) - eA_\LCperp(x^\LCp_i(\pi_\LCm)) \;.
\ee
(Again suppressing dependence of $\pi_\LCperp$ on $x^\LCp$.) It is clear from the integral expressions (\ref{TheFields}) that the results (\ref{HiddenInt0})--(\ref{HiddenInt2}) simply describe the energy transferred to the positron from the background fields over the elapsed time.  This is illustrated in Fig.~\ref{ParBildFig}.

With these definitions we can give a very simple expression for $\mathbb{P}$:
\be\label{Finemang}
	\mathbb{P} = \exp \bigg[-\frac{\pi m^2 + \pi \big[p_\LCperp + \pi_\LCperp(-p_\LCm) \big]^2}{|e|E^{\scriptscriptstyle\parallel}\big(x^\LCp_i(-p_\LCm)\big)} \bigg]\;.
\ee
where $x^\LCp_i$ and $\pi_\LCperp$ are defined in (\ref{HiddenInt1}) and (\ref{HiddenInt2}). We recognise a similar structure as found in Schwinger's results, but for more general fields, and also depending instantaneously on lightfront time. Comparing (\ref{result}) and (\ref{MixedWigner}),  $\mathbb{P}$ is naturally interpreted as the probability that the positron states with momentum $\pi_\LCm=-p_\LCm$ have been filled by time $x^\LCp$, since particles are being created by the background fields.  Using the dynamics discussed above, a more precise statement is the following: $\mathbb{P}$ gives the probability of observing positrons with momenta $\pi_\LCm$ and $\pi_\LCperp$ at time $x^\LCp$, such that these positrons were created at $x^\LCp_i$ with $\pi_\LCm=0$ and transverse momentum normally distributed about $\pi_\LCperp=0$. Note that because the argument of the pair creating field in $\mathbb{P}$ is {\it not}  $x^\LCp$ but $x_i^\LCp$, the probability of observing a positron is dependent on the electric field strength {\it at the moment of creation} $x^\LCp_i$, and not on the `observation' time $x^\LCp$, see also Fig.~\ref{ParBildFig}. This is a neat and physically sensible result.

We now explain this interpretation in more detail and reinforce it with a series of examples. In order to keep the presentation as clear as possible, we begin by dropping down to $1+1$ dimensions, turning off all transverse dependence. This will be reinstated below. The first obvious question, and obvious difference between (\ref{result}) and (\ref{MixedWigner}), is why do we not see electron states being filled?  Why is there no change to the DHW function for $p_\LCm>0$? 

A related result was found and explained in \cite{Tomaras:2000ag,Tomaras:2001vs}, which we now describe in our language. Our DHW function has the form of a disturbance propagating out from $p_\LCm=0$ as time evolves. This is also where the singularities in the Dirac equation live in the $L\to\infty$ limit, and we have already seen that $\mathbb{P}$ describes particles with zero initial longitudinal momentum. Put together, this means that pairs are created travelling at the speed of light. The distinction is that the electrons, being accelerated {\it down} the $x^3$-axis by the positive field, travel parallel to the $x^\LCm$ direction and so, from the perspective of the infinitely boosted lightfront frame, immediately vanish.  The positrons, on the other hand, are accelerated {\it up} the $x^3$-axis and therefore acquire positive $\pi_\LCm$, see also (\ref{HiddenInt0}), and remain visible in the lightfront frame.  We therefore confirm the result of \cite{Tomaras:2000ag,Tomaras:2001vs}: from the lightfront perspective, one only sees the positrons.

The final piece of (\ref{result}) to consider is the second step function. This states simply that the argument of $E^{\scriptscriptstyle\parallel}$, that is $X_{eA(x^\LCp)-\pi_\LCm}$, must be positive; in other words, sufficient time must have elapsed for a particle with zero longitudinal momentum to have acquired $\pi_\LCm$ by time $x^\LCp$. Together, the two theta functions therefore imply, using (\ref{HiddenInt0})-(\ref{HiddenInt2}), that
\be
	0 < x_i^\LCp < x^\LCp \;,
\ee
which is a simple statement of causality: observed pairs must have been created at earlier times, but after the longitudinal field turns on. We conclude that the DHW function shows us pair production from the vacuum, in real lightfront time, with $\mathbb{P}$ the probability of pair creation. The positrons appearing at time $x^\LCp$ with a finite range of momenta $\pi_\LCm$ are subject to the (natural) constraint that sufficient time must have elapsed for the positron to have absorbed this momenta from the fields, starting from $\pi_\LCm=0$.  We now move on to explicit examples, staying in $1+1$ dimensions for the moment.

\subsection{Example: finite pulse duration}
\begin{figure}[t]
\includegraphics[width=0.9\columnwidth]{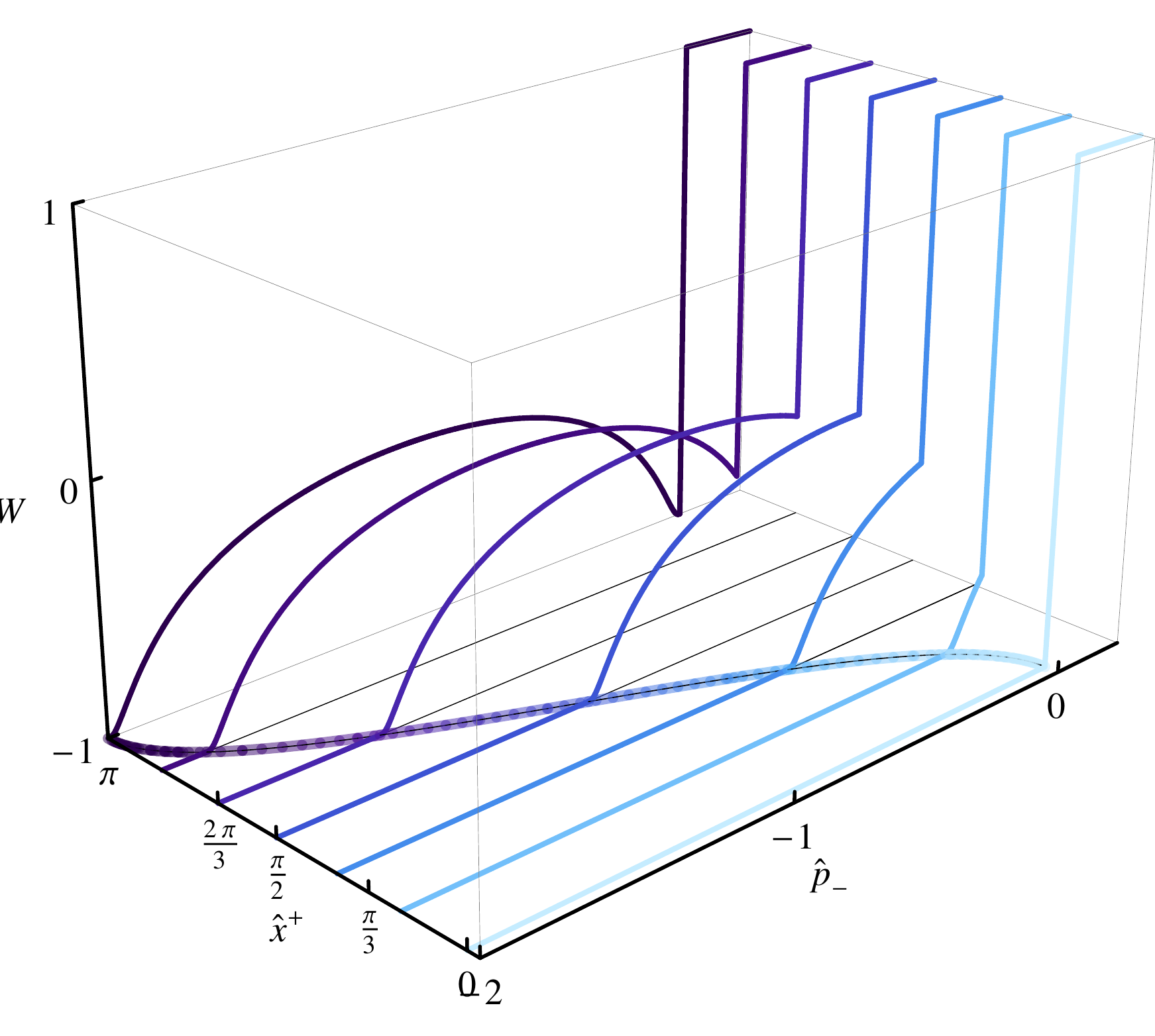}
\caption{\label{FIG:SIN} The DHW function in the subcycle pulse (\ref{pulse1}), plotted as a function of $p_\LCm$, for zero transverse momentum. The maximum allowed momentum, see (\ref{Constraint}) is also shown in the $(\hat{p}_\LCm,\hat{x}^\LCp)$ plane.}
\end{figure}

We begin with the electric field
\be\label{pulse1}
	E^{\scriptscriptstyle \parallel}(x^\LCp) = E_0 \sin(\omega x^\LCp)
\ee
for $0\leq \omega x^\LCp \leq \pi$ and zero otherwise, modelling a half-cycle of the laser. The DHW function $W$ is plotted in Fig.~\ref{FIG:SIN}. When plotting, we use rescaled variables
\be\label{rescaled}
	\epsilon_0\equiv \frac{|e|E_0}{m^2} \;,\quad \hat{x}^\LCp \equiv \omega x^\LCp \;,\quad \hat{p}_\LCm = \frac{p_\LCm\omega}{|e|E_0} \;,
\ee
which measures the electric field strength in units of the Schwinger field and $p_\LCm$ in units of (as we are about to see) half its maximum value. We have chosen $\epsilon_0=\pi$ for our plot, which means our electric field strength is roughly three times higher than the Schwinger limit: this compensates fully for the damping factor in the exponent of (\ref{Finemang}) and allows us to clearly see the behaviour of the DHW function. We consider other field strengths below.

In Fig.~\ref{FIG:SIN}, we see that $W$ matches the free theory result (\ref{Sign}) at $x^\LCp=0$.  As time evolves, $W$ becomes both $p_\LCm$ and $x^\LCp$ dependent, with the deviation from vacuum  spreading out from $p_\LCm=0$ at $x^\LCp=0$. At time $x^\LCp$ the function is explicitly limited in extent by the theta-functions in (\ref{result}), which give
\be\label{Constraint}
	eA_\LCm(x^\LCp) + p_\LCm >0 \iff x^\LCp > X_{-p_\LCm} \;,
\ee
The DHW function eventually stabilises as the field switches off at $x^\LCp=x^\LCp_f$, upon which it is straightforward to extract properties of the final positron distribution. The final range of possible positron momenta is dictated by, following the above, $-p_\LCm>0$ and $-p_\LCm< eA_\LCm(x_f^\LCp)$. In our current example, $\omega x^\LCp_f \equiv \pi$, so the final range of positron momenta is
\be\label{FinalRange}
	0  < \pi_\LCm < 2\frac{|e|E_0}{\omega} \equiv 2ma_0 \;,
\ee
where, in the final equality, we have introduced the peak field intensity $a_0$ \cite{Heinzl:2008rh},
\be
	a_0 = \frac{|e| E_\text{max}}{\omega m}\,.
\ee
The momentum distribution is peaked around that value of $\pi_\LCm$ such that the electric field is maximal {\it at the instant of creation}. Let $x^\LCp_0$ be the time at which $E^{\scriptscriptstyle\parallel}(x_0^\LCp)$ is maximal, then the most probable kinetic momentum $\langle\pi_\LCm\rangle$ is the solution of the equation 
\be
	x_0^\LCp = X_{-\langle\pi_\LCm\rangle + eA_\LCm(x^\LCp_f)} \;.
\ee
For our current example $\omega x_0^\LCp=\pi/2$ and we find 
\be\label{intensity1}
	\langle\pi_\LCm\rangle= \frac{|e|E_0}{\omega} \equiv ma_0 \;,
\ee
which is also clearly seen from Fig.~\ref{FIG:SIN}. Since the momenta are on-shell in the free theory, we have $\langle\pi_\LCp\rangle = m^2/2\langle\pi_\LCm\rangle$ and so we can easily convert these expressions back to cartesian co-ordinates to find the likely energy and $z$-component of the momentum, which are
\be \label{intensity2}
	\sqrt{2}\langle\pi^0\rangle =  \frac{m}{2a_0} + ma_0\;, \quad \sqrt{2}\langle\pi^3\rangle = \frac{m}{2a_0}-ma_0 \,.
\ee
(Note that the probability $\mathbb P$ for producing very high energy particles $\pi_\LCm\simeq 0$ in the range (\ref{FinalRange}) is almost zero.) 
\begin{figure}[t!]
\includegraphics[width=\columnwidth]{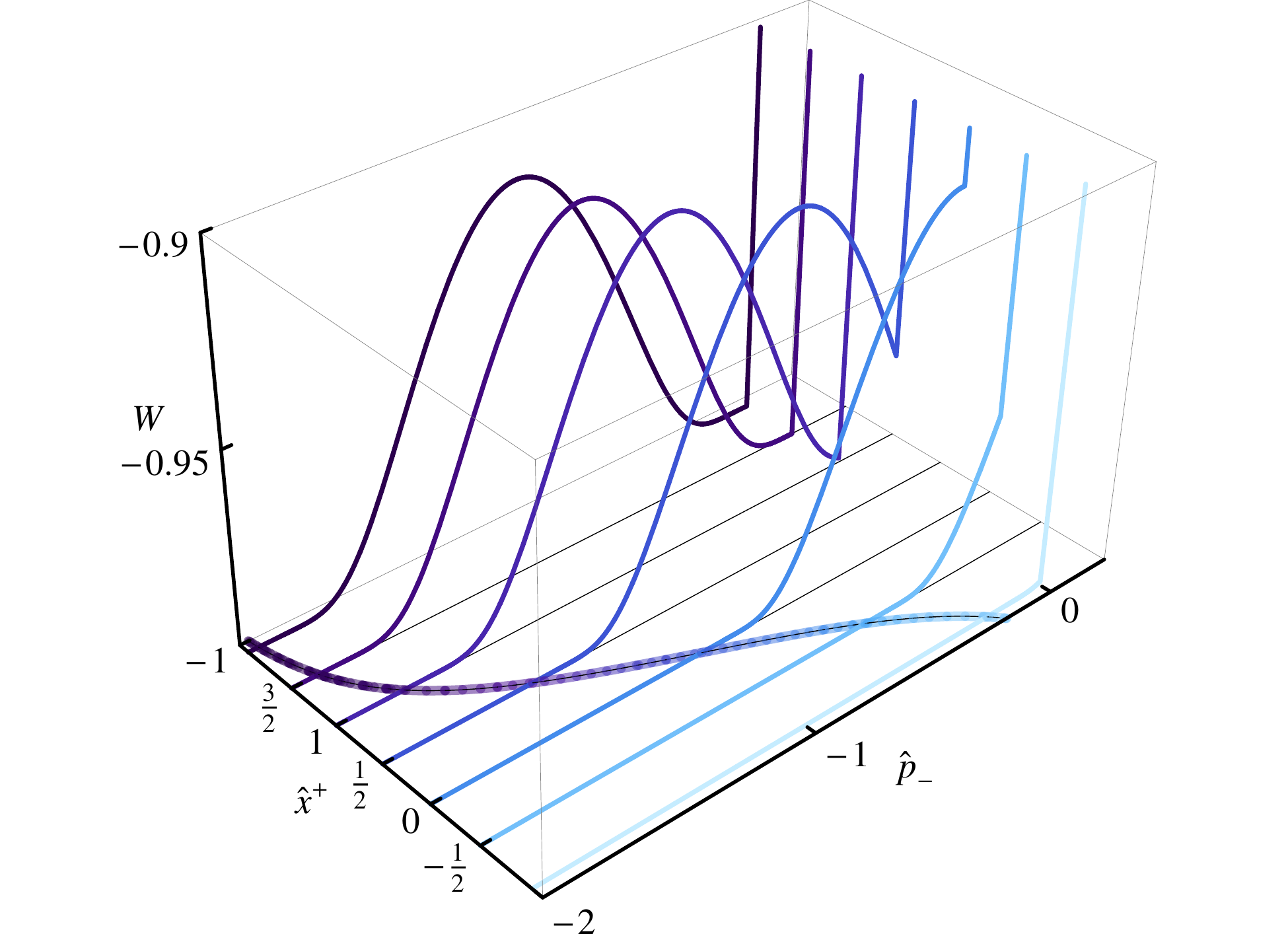}
\caption{\label{FIG:IS} The DHW function in the $\text{sech}^2$ pulse (\ref{pulse2}) with peak amplitude $\epsilon_0=1$, see (\ref{rescaled}), plotted as a function of $p_\LCm$, zero transverse momentum. As before, the range of allowed momentum, see (\ref{Constraint}) is shown in the $(\hat{p}_\LCm,\hat{x}^\LCp)$ plane: there is a smoother falloff than in the previous example.}
\end{figure}
\subsection{Example: adiabatic switching}
In the light of recent literature results, which we discuss below, it is worthwhile pointing out that there is nothing to stop us turning our fields on arbitrarily smoothly starting from arbitrary initial times, without affecting the essential properties of our solutions (\ref{this}) or our DHW function (\ref{result}): our choice of switching the fields on at $x^\LCp=0$ was for convenience.  We therefore consider a field which, while qualitatively similar to our previous example, falls off quickly but smoothly at $\pm\infty$, namely
\be\label{pulse2}
	E^{\scriptscriptstyle\parallel}(x^\LCp) = E_0\, \text{sech}^2(\omega x^\LCp) \;,
\ee
where $E_0$ gives the peak intensity and $1/\omega$ the effective duration of the pulse. The corresponding gauge potential is given by the integral from $x^\LCp=-\infty$ of this function, and is therefore
\be
	eA_\LCm(x^\LCp) = \frac{|e| E_0}{\omega} \big(1+\tanh(\omega x^\LCp)\big) \;,
\ee
where the constant term follows from the definition (\ref{TheFields}), with the initial time translated to $x^\LCp=-\infty$. The resulting DHW function is plotted in Fig.\ \ref{FIG:IS} for a peak field strength equal to the Schwinger field. The DHW function is shown over the temporal range $-1/2<\omega x^\LCp<2$. Before this, the deviation from vacuum is minimal and afterwards the DHW function becomes effectively stable.  The final range of positron momenta which may be created by the field remains finite and, with our choice of variables, is expressed just as before, $0<\pi_\LCm < 2m a_0$. The probability $\mathbb{P}$ is peaked, in the limit $x^\LCp\to\infty$, around $\pi_\LCm = ma_0$.  Note that even at the Schwinger limit, the probability of pair production remains small.
\subsection{Example: ever-increasing field}
\begin{figure}[t]
\includegraphics[width=0.9\columnwidth]{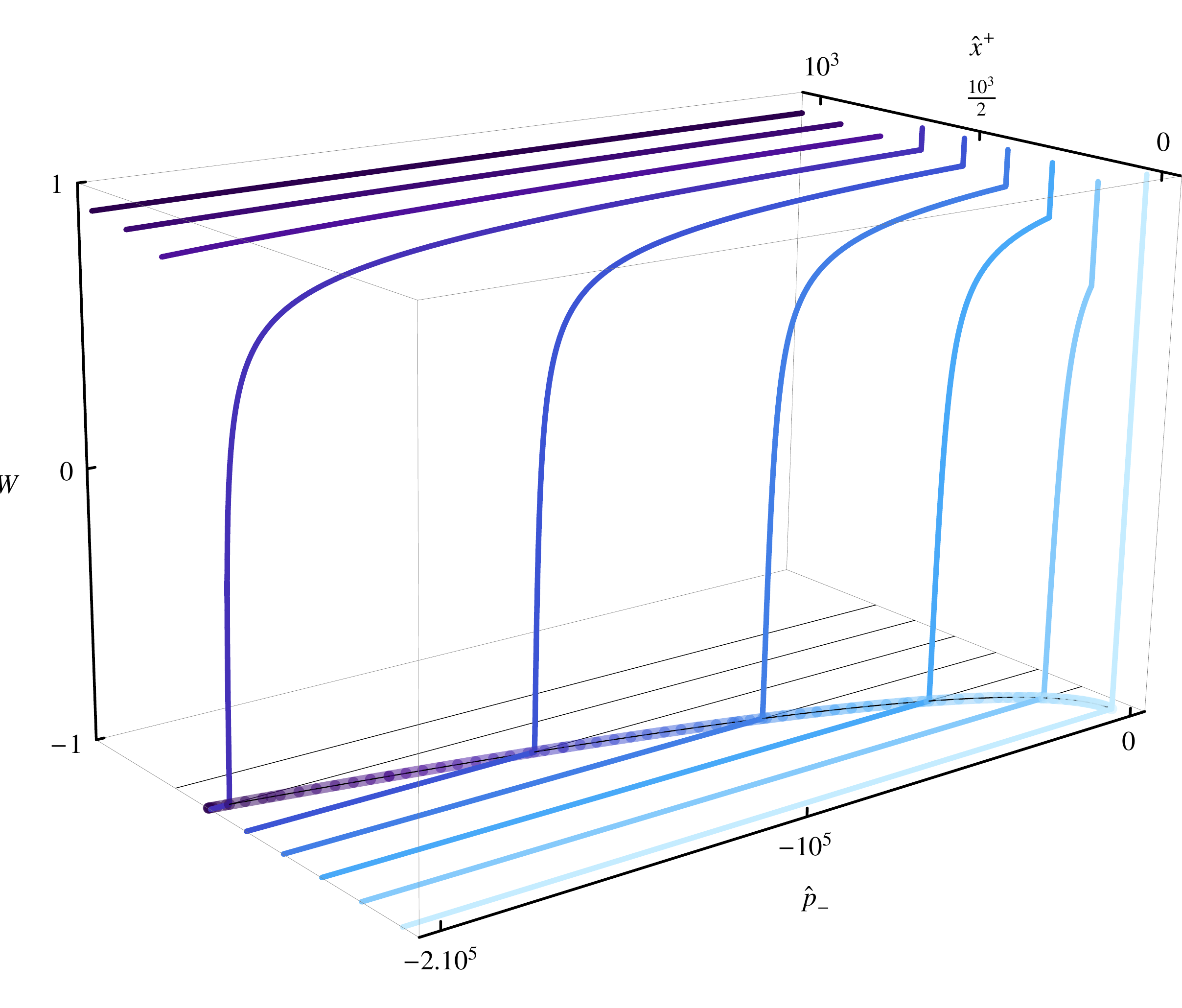}
\caption{\label{newfig} DHW function in the field (\ref{wrong}), with $\epsilon_0=0.5$, which increases without bound. As time evolves, all positron states are filled with unit probability.}
\end{figure}
We note that the factor of $2$ in (\ref{result}) can be understood from a second perspective. Suppose we consider an electric field which increases without bound, for example
\be\label{wrong}
	E^{\scriptscriptstyle\parallel}(x^\LCp) = E_0 \omega x^\LCp\;.
\ee
As time increases, the electric field becomes overcritical and we expect the probability of creating particles with any given momentum (in the allowed range, which also expands in time) to approach unity. This means that the state should become filled with positrons, and we expect to recover, from (\ref{EitherFilled}), $W=+1$ in the region $p_\LCm<0$. This is precisely what the factor of $2$ ensures: if $\mathbb{P}\to 1$, the DHW function approaches $W=1$ for $p_\LCm<0$. This is shown explicitly in Fig.~\ref{newfig}: the DHW function transfers from $-1$ to $+1$ over the whole negative $p_\LCm$ range as time evolves.
\subsection{Comparison with the instant-form approach} \label{TalkSect}
\begin{figure}[t!]
\includegraphics[width=0.85\columnwidth]{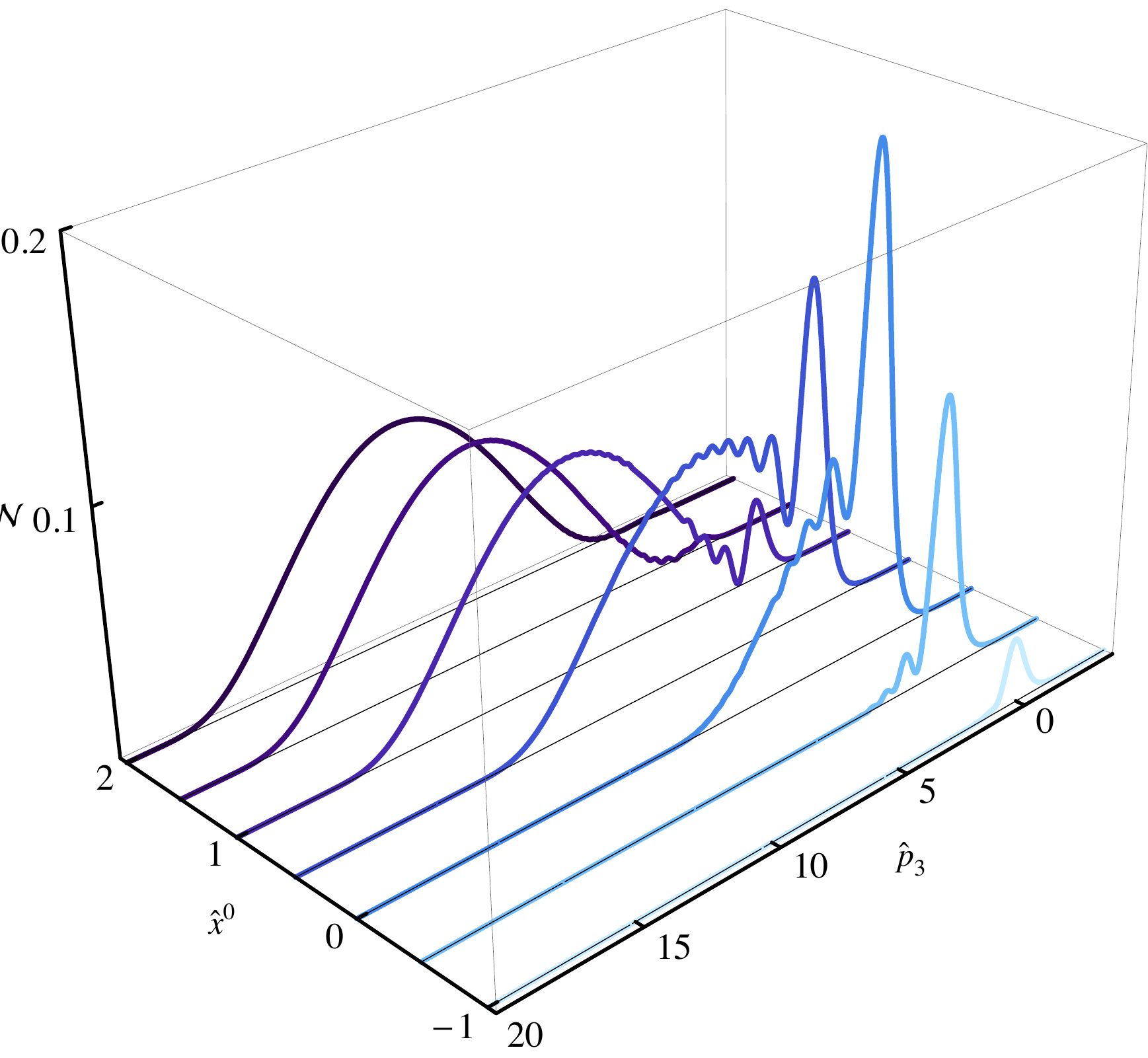}
\caption{\label{FIG:INST} Particle number density $\mathcal{N}$ for the $x^0$-dependent $\text{sech}^2$ pulse within the instant--form DHW formalism, plotted as a function of $\hat{p}_3$ and zero transverse momentum. Peak amplitude $\epsilon_0=1$.}
\end{figure}

Our results share some similarities with investigations of particle creation in $x^0$--dependent electric fields, within the usual (instant-form) DHW formalism. In that approach, a certain combination of instant-form DHW spinor components can be interpreted as a particle number density \cite{Hebenstreit:2011pm}, and its behaviour is as follows. As the electric field grows with time $x^0$, energy is transferred to the Dirac field such that a peak around $p_3=0$ develops. At intermediate (non-asymptotic) times this peak is interpreted as being composed of virtual electron-positron pairs. As time evolves, only a part of these virtual particles become real particles which are then accelerated by the electric field and spread out from $p_3=0$. At asymptotic times $x^0\to\infty$, the particle number density of real particles stabilises whereas the virtual electron-positron peak around $p_3=0$ disappears again. An example is shown in Fig.~\ref{FIG:INST} for the asymptotically switched field $E(x^0)= E_0\operatorname{sech}^2(\omega x^0)$.  For more details see \cite{Hebenstreit:2011pm}.

One difference between these results and our own is that the lightfront DHW function does not exhibit the intermediate virtual particle peak or oscillatory structure seen in the instant form. The reason for this seems to be that pair production on the lightfront is an instantaneous event, occurring at the instant when a given fermion mode can produce a particle of zero longitudinal momentum, see (\ref{HelloZero}).  This is confirmed by our expression for the pair creation probability $\mathbb P$: it is expressed entirely in terms of classical particle trajectories.

This is quite intriguing, as it may be related to the `triviality' of the lightfront vacuum. The instant--form vacuum is filled with virtual pairs which can be pulled onto the mass--shell by the external field. Recall that the Schwinger field strength can be obtained by equating the electron rest mass with the work done by the electric field over the lifetime of a virtual pair: in this sense, there is a time scale involved in Schwinger pair production. The lightfront vacuum, on the other hand, is often referred to as `trivial', which is the statement that it is completely empty of particles, both real and virtual \cite{Heinzl:2003jy}. Moreover, it is stable. In this picture, then, pairs are created from the energy pumped into the system, not from virtual particles being pulled on-shell, and the Schwinger `time-scale' is absent. This is an investigation for another time, though. We now return to properties of the DHW function.

\subsection{Nonperturbative dependence}
Recent results on pair production in $x^0$-dependent electric fields (in the usual instant time DHW formalism), find a purely perturbative dependence on a particular electric field which switches on adiabatically in the infinite past \cite{BialynickiBirula:2011eg}. Moreover, it is stated that the essential singularity of Schwinger's results must therefore be due to the unphysical nature of a constant, ever-present electric field.

Let us reconsider our results in this light.  We have seen explicitly that for both sharply and smoothly switched fields, the pair production probability is basically described by a factor $\exp(-m\pi/|e| E^{\scriptscriptstyle\parallel})$, which retains the essential singularity in the coupling from Schwinger's result. It is therefore clear that the nonperturbative nature of pair creation is not due to some unphysical assumption about when, or how smoothly, the fields turn on or off. The dependence of our results on $e$ is of course a little more complex than that, since $E^{\scriptscriptstyle\parallel}$ is evaluated at $x^\LCp_i(-p_\LCm)$, see (\ref{Finemang}).
 
It is useful to examine the form of the (final) probability $\mathbb{P}$ when the fields turn off. We work with the physical momentum $\pi_\LCm$ here, and turn off the plane wave fields for simplicity, setting also $\pi_\LCperp=0$. We begin with the field (\ref{pulse1}), for which the final probability is
\be
\begin{split}
	-\log\mathbb{P}\big|_{\omega x^\LCp=\pi} &= \frac{\pi m^2/\omega}{\sqrt{\pi_\LCm(2ma_0-\pi_\LCm)}} \;.
	\end{split}
\ee
Note that the denominator is positive because of the finite allowed $\pi_\LCm$ range. We can examine this probability for, for example, small momenta (which corresponds to extremely energetic particles in the lab frame)
by expanding in $\pi_\LCm$:
\be
\begin{split}
	 \frac{\pi m^2}{\sqrt{2|e|E_0\omega\pi_\LCm}}\bigg(1 + \frac{\omega\pi_\LCm}{4|e|E_0}+\ldots\bigg)
\end{split}
\ee
We clearly see the $1/|e|E_0$ dependence. What happens for the adiabatically switched $\text{sech}^2$ pulse? For this field, the final probability is
\be
	\lim_{x^\LCp\to\infty} -\log\mathbb{P} = \frac{\pi|e|E_0m^2}{2|e|\omega E_0 \pi_\LCm-\omega^2\pi_\LCm^2} \;.
\ee
Again, the denominator is positive, and we can make the same small $\pi_\LCm$ expansion as above, finding
\be\label{was}
	\lim_{x^\LCp\to\infty} -\log\mathbb{P} \simeq \frac{\pi m^2}{2\omega\pi_\LCm} \bigg(1 + \frac{\omega\pi_\LCm}{2|e|E_0}+\ldots\bigg)\;.
\ee
This displays a different dependence on the various parameters. In particular, the dominant term is independent of the field strength $E_0$. Does this correspond to a perturbative dependence on the field strength? The answer is no: not only does the second term in the expansion contain explicitly nonperturbative (Schwinger-like) terms, but the leading term of (\ref{was}) actually contains a hidden nonperturbative dependence. To see this, note that the leading term, despite not being explicitly dependent on $E_0$, {\it does not} survive the limit $E_0\to 0$ (which would contradict the free theory). This is because the range of $\pi_\LCm$ is finite, being limited by $eA_\LCm$, so that taking $E_0\to 0$ implies taking $\pi_\LCm\to 0$, and this kills $\mathbb{P}$ as the fields turn off.

What this result really shows is only that,  and as is not surprising, the distribution of the produced particles depends on the geometry of the field, for example whether the field turns off sharply or smoothly. We have not been able to identify a regime where the results may be expressed as a perturbation in $E_0$.  We stress that this holds at least on the lightfront, for fields depending on $x^\LCp$: there are differences between this and the instant-form approach, see Sect.~\ref{TalkSect}, above.

It is, though, entirely possible for the effective action to exhibit {\it both} perturbative and Schwinger-like nonperturbative behaviour when the electric field depends on $x^0$ (and is even adiabatically switched), depending on the relative sizes of the parameters involved. This is shown in \cite{Dunne:2002rq}, which discusses many deep connections between perturbative and nonperturbative physics. One is lead to conclude that the perturbative dependence found in \cite{BialynickiBirula:2011eg}, while very interesting, is not inconsistent with the existence of nonperturbative behaviour. 
Combining this with our own results, we do not believe that any doubt is cast on the validity of Schwinger's result.

\subsection{Transverse dependence}
Finally, it is time to return to $3+1$ dimensions proper, and allow for plane wave fields. Consider the full expression (\ref{Finemang}) for the probability. The plane wave contributions do not appear in the step functions, thus they do not affect the constraints dictating the momentum ranges.  Nor do the plane waves enter into the argument of the pair creating field $E^{\scriptscriptstyle\parallel}$.

It is clear that without the plane waves, the probability for production of pairs with nonzero transverse momenta is normally distributed around $\pi_\LCperp=0$. Turning on the plane waves, it may seem strange at first glance that the peak of this distribution is shifted to nonzero values: is the plane wave affecting the probability of pair production? The answer is no: recalling (\ref{HiddenInt0})-(\ref{HiddenInt2}), the distribution in (\ref{Finemang}) is obtained for positrons {\it created} with transverse momentum normally distributed around $\pi_\LCperp=0$, and which at the subsequent time $x^\LCp$ must have acquired transverse momentum
\be
	\int\limits^{x^\LCp}_{x^i(\pi_\LCm)} \!\ud s\ \sqrt{2}e E^\LCperp(s)\equiv \pi_\LCperp(\pi_\LCm) \;,
\ee
from the plane wave fields, using the Lorentz equation. The DHW function therefore takes into account both what happens at the instant of creation, but also what would subsequently be observed.

To summarise, the plane waves do of course influence the particles after they are created, and so it is no surprise that they appear in the DHW function and the final particle distribution. The plane waves do have no influence, though, on whether particles are created or not. This is reaffirmed by integrating over momenta to obtain, for example, the total probability of pair production or the vacuum persistence amplitude: one finds that all dependence on the plane wave fields vanishes because $\pi_\LCperp(-p_\LCm)$ can be absorbed into $p_\LCperp$ by a change of variable. Thus the plane wave fields, and in particular any effective mass they may generate, do not influence the probability of vacuum decay \cite{Narozhnyi:1976zs,Tomaras:2000ag,Tomaras:2001vs}.   

\section{Conclusions}\label{ConcSect}
We have investigated the phenomenon of non-perturbative pair creation in background electromagnetic fields, within the lightfront DHW formalism. We calculated the DHW function by solving the Dirac equation in a combination of longitudinal electric and transverse plane wave fields which both depend arbitrarily on lightfront time. This extends the work of \cite{Nikishov:1964zza,Tomaras:2000ag,Tomaras:2001vs} to an even wider class of fields.

As shown in \cite{Hebenstreit:2010cc}, the DHW function $W$ is not altered by a single plane wave field, since a plane wave can not produce pairs. Switching on an additional pair-creating electric field, however, one observes  a deviation from the vacuum result which signals pair creation. The pair creation probability itself is exponentially suppressed by a factor $\exp(-m\pi/|e| E^{\scriptscriptstyle\parallel})$ which retains the essential singularity in the electromagnetic coupling $e$ in Schwinger's result, but is valid for much more general fields, and in particular is independent of how the electric field is switched on and off. This may be contrasted with recent results in the instant--form approach \cite{BialynickiBirula:2011eg}.

Notably, we have seen that the value of the DHW function $W$ is altered only in its positron sector, whereas the electron content remains unchanged. This confirms previous results that, from the lightfront perspective, only one of the particle species remains in the theory following creation \cite{Tomaras:2000ag,Tomaras:2001vs}.

All in all, the DHW function can be a powerful tool in analysing quantum physics in background fields (particularly pair production), and in a language which is essentially classical. We have given an elegant and physical interpretation in terms of the pair creation probability and the subsequent dynamics of the particles. This makes it clear that we observe real particle production in the lightfront formalism.\\

\acknowledgments
A.~I.\ gratefully thanks Richard Woodard for correspondence, and Tom Heinzl for discussions.  Fig.\ 1 to Fig.\ 4 created using JaxoDraw \cite{Binosi:2003yf,Binosi:2008ig}.

F.~H.\  is supported by the Baltic Foundations. A.~I.\ and M.~M.\ are supported by the European Research Council Contract number 204059-QPQV. This work was performed under the \emph{Light in Science and Technology} Strong Research Environment, Ume{\aa} University.

\appendix
\section{Notation.}
Our lightcone directions are $x^\LCpm = (x^0 + x^3)/\sqrt{2}$. We prefer momenta to carry covariant indices and so $p_\LCm$ is a spatial momentum conjugate to $x^\LCm$, while $p_\LCp$ is the lightfront energy, conjugate to lightfront time $x^\LCp$. The metric has determinant $-1$ as in Cartesian coordinates and contra-/covariant indices are related by $v^\LCmp = v_\LCpm$ for arbitrary vectors $v$.  We use a sans-serif font to denote the `spatial' variables and momenta, i.e.\ ${\sf x}\equiv \{x^\LCm,x^\LCperp\}$ and ${\sf p}\equiv \{p_\LCm,p_\LCperp\}$. Our Fourier conventions are
\be
	f(p) = \int\!\ud x\ e^{ipx} f(x) \;, \quad 	f(x) = \int\!\frac{\ud p}{2\pi}\ e^{-ipx} f(p) \;,
\ee
and we Fourier transform before taking conjugates so
\be
	f^\dagger(p) = \int\!\ud x\ e^{-ipx} f^\dagger(x) \;, \quad 	f^\dagger(x) = \int\!\frac{\ud p}{2\pi}\ e^{ipx} f^\dagger(p) \;.
\ee
%

\begin{widetext}

\section{The quantum theory}
We now wish to quantise our solution (\ref{this}). Quantisation is performed by imposing canonical commutation relations on the initial data. As in \cite{Tomaras:2000ag,Tomaras:2001vs} one finds that the quantised spinor fields obey the desired commutation relations, returning briefly to full co-ordinate space,
\be\label{BulkCCRs}
	\big\{\psi_\LCp(x),\psi_\LCp^\dagger(y)\big\}\big|_{x^\LCp=y^\LCp} = \frac{1}{\sqrt{2}}\Lambda^\LCp \delta^{2}(x^\LCperp-y^\LCperp)\delta(x^\LCm-y^\LCm) \;,
\ee
provided that the initial data obeys
\be\label{InitialCCRs}\begin{split}
	\big\{\psi_\LCp(x),\psi_\LCp^\dagger(y)\big\}\big|_{x^\LCp=y^\LCp=0} &= \frac{1}{\sqrt{2}}\Lambda^\LCp\delta^{2}(x^\LCperp-y^\LCperp)\delta(x^\LCm-y^\LCm)\;, \\
	\big\{\psi_\LCm(x),\psi_\LCm^\dagger(y)\big\}\big|_{x^\LCm=y^\LCm=-L} &= \frac{1}{\sqrt{2}}\Lambda^\LCm\delta^{2}(x^\LCperp-y^\LCperp)\delta(x^\LCp-y^\LCp)\;,
\end{split}
\ee
What we are really doing here is solving a Cauchy problem with initial data on two lightlike characteristics: this is not quite what one usually does on the lightfront (where the operators on the $x^\LCm$ characteristic are not needed explicitly) but is necessitated by the time-dependence introduced into the zero-mode problem.  Note that without the operators on $x^\LCm=-L$ one does not recover known results such as Schwinger's vacuum persistence amplitude in the $L\to\infty$ limit. Moreover, the approach used here has been verified by alternative methods \cite{Fried:2001ur}. (The question of whether there is a method to recover the results of this approach in ordinary lightfront quantisation without the operators at $x^\LCm=-L$ has not, to our knowledge, been addressed.)

From here one can construct the Hamiltonian and states of the theory as normal. Computationally, it is useful to work in what becomes Fourier space in the $L\to\infty$ limit: considering the $k_\LCm$-integrations inside the functions $G$, (\ref{G-DEF}), we  define the `almost' Fourier transform $\tilde\psi_\LCp$ by 
\be
	\psi_\LCp(x^\LCp,x^\LCm) \equiv \int\!\frac{\ud k_\LCm}{2\pi}\ e^{-i x^\LCm(k_\LCm+i/L)}\tilde\psi_\LCp(x^\LCp,k_\LCm) \;.
\ee
Dependence on $k_\LCperp$ will not be written unless it is needed. Explicitly, our new field is 
\be\label{this2}
	\tilde\psi_\LCp(x^\LCp,k_\LCm) = \int\limits_{-L}^\infty\!\ud y^\LCm e^{iy^\LCm(k_\LCm+i/L)} \mathcal{E}_{k_\LCm}(0,x^\LCp)\psi_\LCp(0,y^\LCm)  +\int\limits_0^{x^\LCp}\!\ud y^\LCp \frac{e^{-iL(k_\LCm+i/L)}}{k_\LCm-eA_\LCm(y^\LCp)+i/L}  \mathcal{E}_{k_\LCm}(y^\LCp,x^\LCp) i\partial_\LCp\psi_\LCp(y^\LCp,{\scriptstyle -L})\;.
\ee
(The operators $\tilde\psi_\LCp$ become canonically normalised creation (and annihilation) operators as $L\to\infty$, in the interpretation of \cite{Tomaras:2000ag,Tomaras:2001vs}, up to a factor of $2^{1/4}$: hence the factor of $\sqrt{2}$ in our DHW function (\ref{defn}).) The state of interest for us is the vacuum, or rather the {\it free} vacuum, since what we wish to do is begin in this state at $x^\LCp=0$, apply our external fields and see what happens as time evolves. Since we are working in the Heisenberg picture, in which the states are time-independent, the state we need is precisely the free vacuum state. This means that the density we need, (\ref{Udef}), is calculated by first expressing $\psi_\LCp$ in terms of the free initial data using (\ref{this2}), and then evaluating the density of these free fields in the free vacuum state. This calculation is performed using the ordinary free-field mode expansion. For example, a free-field calculation easily gives us
\be\label{FreeFieldRes}
	\bra{0} \psi_\LCp(x) \psi^\dagger_\LCp(y)\ket{0}\big|_{x^\LCp=y^\LCp=0} =
	\frac{1}{\sqrt{2}}\Lambda^\LCp\int\limits_0^\infty\!\frac{\ud k_\LCm}{2\pi} e^{-ik_\LCm (x_2-x_1)^\LCm}  \delta^2(x^\LCperp-y^\LCperp)\;,
\ee
which we will use below.

\section{Calculating the DHW function}
We now wish to calculate (\ref{Udef}) and (\ref{defn}). To do so we make an assumption on the longitudinal electric field, namely that it is positive, e.g. modelling a subcycle pulse. As a result, $eA_\LCm$ is a positive function increasing from $0$ at $x^\LCp=0$. This approximation can be relaxed, but doing so means that the analytic structure in the functions $\mathcal{E}$ becomes significantly more complex, since the kinematic momentum $\pi_\LCm(x^\LCp) = k_\LCm-eA_\LCm(x^\LCp)$ may have multiple zeros. The `subcycle assumption' means that this function has at most one zero for a given $k_\LCm$. With this assumption, we can define the inverse function $X(k_\LCm)\equiv X_{k_\LCm}$ by
\be\label{X}
	k_\LCm - eA_\LCm(X_{k_\LCm})\equiv 0\;.
\ee
This function exists for the duration of the longitudinal field $E^{\scriptscriptstyle\parallel}$. What happens after the fields turn off is explained below. To proceed, we express the density $U$ in terms of the modes $\tilde\psi_\LCp(x^\LCp,k_\LCm)$. The first term of the commutator is
\be\label{start}
	\bra{0}\psi_\LCp(x^\LCp,x^\LCm_2)\psi^\dagger_\LCp(x^\LCp,x^\LCm_1)\ket{0} = \int\!\frac{\ud l_\LCm}{2\pi} f(l_\LCm) e^{-i(l_\LCm+\tfrac{i}{L})x_2^\LCm} \int\!\frac{\ud q_\LCm}{2\pi}f^*(q_\LCm) e^{i(q_\LCm-\tfrac{i}{L})x_1^\LCm} \bra{0}\tilde\psi_\LCp(x^\LCp,l_\LCm){\tilde\psi}^\dagger_\LCp(x^\LCp,q_\LCm)\ket{0} \;,
\ee
where we have included test functions under the momentum integrals in order to take the $L\to\infty$ limit more rigourously. Since $\psi_\LCp$ itself contains two terms, the overlap in (\ref{start}) contains four. Only one of these, the product of the first term in (\ref{this2}) with its conjugate, ultimately contributes, so we present only this calculation. Using (\ref{this2}) to write down this term, and the free-field expression (\ref{FreeFieldRes}), one finds that the overlap in (\ref{start}) is
\be\begin{split}
 &\frac{1}{\sqrt{2}}\Lambda^\LCp\int\limits_0^\infty\!\frac{\ud k_\LCm}{2\pi}\int\limits_{-L}^\infty\! \ud v\, e^{i(l_\LCm+i/L)v}\int\limits_{-L}^\infty\! \ud z\, e^{-i(q_\LCm-i/L)v}\mathcal{E}_{l_\LCm}(0,x^\LCp)\mathcal{E}^*_{q_\LCm}(0,x^\LCp)e^{-ik_\LCm(v-z)} \\
	&=\frac{1}{\sqrt{2}}\Lambda^\LCp\int\limits_0^\infty\!\frac{\ud k_\LCm}{2\pi}\, e^{-iL(l_\LCm+i/L)} e^{iL(q_\LCm-i/L)}\frac{\mathcal{E}_{l_\LCm}(0,x^\LCp)}{l_\LCm-k_\LCm+i/L}\ \frac{\mathcal{E}^*_{q_\LCm}(0,x^\LCp)}{q_\LCm-k_\LCm-i/L}\;, 
\end{split}\ee
carrying out the $v$ and $z$ integrations to arrive at the second line. We now bring in the momentum integrals and change variables $l_\LCm\to a=L(l_\LCm-k_\LCm)$ and $q_\LCm \to b=L(q_\LCm- k_\LCm)$:
\be\begin{split}\label{huge}
(\text{\ref{start}}) =& \frac{1}{\sqrt{2}}\Lambda^\LCp\int\limits_0^\infty\!\frac{\ud k_\LCm}{2\pi}e^{-ip(x_2^\LCm-x_1^\LCm)}\\
&\bigg[\int\!\frac{\ud a}{2\pi} f(\tfrac{a}{L} + k_\LCm) \frac{\mathcal{E}_{a/L+k_\LCm}(0,x^\LCp)}{a+i}e^{-i(a+i)(1+x^\LCm_2/L)}\bigg] 
\bigg[\int\!\frac{\ud b}{2\pi} f^*(\tfrac{b}{L} + k_\LCm) \frac{\mathcal{E}^*_{b/L+k_\LCm}(0,x^\LCp)}{b-i}e^{i(b-i)(1+x^\LCm_1/L)}\bigg]\;.
\end{split}
\ee
Our task now is to take the $L\to\infty$ limit. The behaviour of the $\mathcal{E}$ functions in this limit depends crucially on the relative values of $x^\LCp$ and $k_\LCm$, as these determine whether or not we hit the singularity. Explicitly, we have
\be\label{scontour}
	\mathcal{E}_{a/L+k_\LCm}(0,x^\LCp) = \exp\bigg[-\frac{i}{2}\int\limits_0^{x^\LCp}\!\frac{\ud s\, \omega^2(s)}{k_\LCm-eA_\LCm(x^\LCp) + \tfrac{a+i}{L}}\bigg] \;.
\ee
We have $k_\LCm>0$. It is therefore clear that if $k_\LCm - eA_\LCm(x^\LCp)>0$ there are no problems taking the $L\to\infty$ limit. Although we cannot carry out the $s$-integral exactly we know that $\mathcal{E}$ becomes a phase, and this will cancel between the two large bracketed terms in (\ref{huge}). We can therefore write {\it part} of the solution immediately:
\be\begin{split}
	\bra{0}\psi_\LCp(x^\LCp,x^\LCm_2)\psi^\dagger_\LCp(x^\LCp,x^\LCm_1)\ket{0} &\to \frac{1}{\sqrt{2}}\Lambda^\LCp_{\alpha\beta}\int\limits_{eA_\LCm(x^\LCp)}^\infty\!\frac{\ud k_\LCm}{2\pi}\; |f(k_\LCm)|^2 e^{-ik_\LCm(x_2^\LCm-x_1^\LCm)}\int\!\frac{\ud a}{2\pi} \frac{e^{-i(a+i)}}{a+i} \int\!\frac{\ud b}{2\pi} \frac{e^{i(b-i)}}{b-i} +\ldots \\
	&=\frac{1}{\sqrt{2}}\Lambda^\LCp_{\alpha\beta}\int\limits_{eA_\LCm(x^\LCp)}^\infty\!\frac{\ud k_\LCm}{2\pi}\; |f(k_\LCm)|^2 e^{-ik_\LCm(x_2^\LCm-x_1^\LCm)}+\ldots
\end{split}
\ee
The integral over the range $k_\LCm\in 0\ldots eA_\LCm(x^\LCp)$ requires more care. In this case there is always a value of $s$ such that the integral in (\ref{scontour}) acquires an imaginary part which does not drop out of (\ref{huge}). To identify the imaginary part we change variables $eA_\LCm(s)\equiv t$, i.e.\ $s=X_t$ and then expand the resulting $t$ dependence in the numerator:
\be
	\mathcal{E}_{a/L+k_\LCm}(0,x^\LCp) =  \exp\bigg[-\frac{i}{2}\int\limits_0^{eA_\LCm(x^\LCp)}\!\frac{\ud t\, \omega^2(X_t)X'_t}{k_\LCm-t + \tfrac{a+i}{L}}\bigg] =  \exp\bigg[-i\lambda(k_\LCm)\int\limits_0^{eA_\LCm(x^\LCp)}\!\frac{\ud t}{k_\LCm-t + \tfrac{a+i}{L}} + \ldots\bigg]  
\ee
where we have shown only the first term in the series and defined the function
\be\label{xi}
	\lambda(p) = \frac{\omega^2(X_p)}{2|e|E^{\scriptscriptstyle\parallel}(X_p)} \;.
\ee
Carrying out the integrals and then taking the limit $L\to\infty$ one finds
\be
\mathcal{E}_{a/L+k_\LCm}(0,x^\LCp) \to \exp\bigg[-i\lambda(k_\LCm)\big(-\log[|1-eA_\LCm(x^\LCp)/k_\LCm|]-i\pi + \text{real}\big)\bigg] \;,
\ee
where the real logarithm and $i\pi$ come from the first term of the Taylor expansion, and all other terms give real, $a$-independent contributions. (This  behaviour is different from that in the second commutator term calculated in \cite{Tomaras:2001vs}, where the singularity always lies precisely at one of the integral's limits, rather than between them, and the $a$ and $b$ integrals are more complex.) Thus, each $\mathcal{E}$ function contributes the exponential of $-\pi\lambda$ and the rather simple, final expression for the first term of the density (\ref{start}) is
\be\label{FirstTerm}
\bra{0}\psi_\LCp(x^\LCp,x^\LCm_2)\psi^\dagger_\LCp(x^\LCp,x^\LCm_1)\ket{0} =  \frac{1}{\sqrt{2}}\Lambda^\LCp{\!\!\!\!}\int\limits_{eA_\LCm(x^\LCp)}^\infty\!\frac{\ud k_\LCm}{2\pi}\; |f(k_\LCm)|^2 e^{-ik_\LCm(x_2^\LCm-x_1^\LCm)}+\int\limits^{eA_\LCm(x^\LCp)}_0\!\frac{\ud k_\LCm}{2\pi} |f(k_\LCm)|^2 e^{-ik_\LCm(x_2^\LCm-x_1^\LCm)} e^{-2\pi\lambda(k_\LCm)}\;.
\ee
Before moving on to the DHW function itself, it is worth checking this result. The second commutator term has been calculated independently in \cite{Tomaras:2001vs}, see equations (4.2) and (4.9) therein. That calculation is in two dimensions, so if we turn off our transverse dependences completely, one finds, {\it adding} their result to our own, (\ref{FirstTerm}), then we correctly obtain the (1+1) field anticommutator (\ref{BulkCCRs}). Given this positive result, we return to four dimensions, and calculate the second term of the density $U$ from (\ref{BulkCCRs}). The final expression for the density is, removing the test functions,
\be
	U = \frac{e^{-ik_\LCperp y^\LCperp}}{\sqrt{2}} \Lambda^\LCp  \bigg[  \int\limits_{-\infty}^\infty\! \frac{\ud k_\LCm}{2\pi}\ \text{Sign}\big(k_\LCm - eA_\LCm(x^\LCp)\big) e^{-ik_\LCm y^\LCm} + 2\int\limits_0^{eA_\LCm(x^\LCp)}\! \frac{\ud k_\LCm}{2\pi}\ e^{-ik_\LCm y^\LCm} e^{-2\pi\lambda}\bigg] \;,
\ee
where we have written $\sf{x}_2\equiv {\sf x}+{\sf y}/2$, $\sf{x}_2\equiv \sf{x}-\sf{y}/2$ as in Sect.~\ref{WigSect}, and $\lambda$ in this expression is
\be
	\lambda = \frac{m^2 + [k_\LCperp-eA_\LCperp(X_{k_\LCm})]^2}{2 |e| E^{\scriptscriptstyle\parallel}( X_{k_\LCm} )} \;.
\ee
We now perform the Fourier transform in (\ref{defn}). Including the Wilson line, the transformation sets
\be
	p_\LCperp + e A_\LCperp(x^\LCp) = k_\LCperp \;,\quad\text{and}\quad p_\LCm + eA_\LCm(x^\LCp) = k_\LCm \;.
\ee
The final expression for the DHW function is, over the duration of the pulse,
\be\label{result2}
	W(x^\LCp;\mathsf{x},\mathsf{p}) = \text{Sign}(p_\LCm) + 2\,\mathbb{P}\,\theta(-p_\LCm)\theta\big(eA_\LCm(x^\LCp)+p_\LCm\big)\;,
\ee
where the full expression for $\mathbb{P}$ is
\be\label{Finemang2}
	\mathbb{P} = \exp \bigg[-\frac{\pi m^2 + \pi \big[p_\LCperp + eA_\LCperp(x^\LCp)-eA_\LCperp(X_{p_\LCm+eA_\LCm(x^\LCp)}))\big]^2}{|e|E^{\scriptscriptstyle\parallel}\big(X_{p_\LCm+eA_\LCm(x^\LCp)})} \bigg]\;.
\ee

This is the expression given in equation (\ref{Finemang}): the more compact and intuitive notation used therein is explained in the text. Note that $\mathbb{P}$ is gauge invariant by definition, and hence $A_\LCm$, for example, is actually the lightfront integral of the electric field, see (\ref{TheFields}). 
\end{widetext}


\begin{thebibliography}{100}

\bibitem{XFEL}
\href{http://xfel.desy.de/}{\texttt{http://xfel.desy.de/}}

\bibitem{ELI}
\href{http://www.extreme-light-infrastructure.eu/}{\texttt{http://www.extreme-light-infrastructure.eu/}}



\bibitem{Dunne:2010zz}
  G.~V.~Dunne,
  Int.\ J.\ Mod.\ Phys.\  A {\bf 25} (2010) 2373.

\bibitem{original} E.\ P.\ Wigner, 
Phys.\ Rev.\ \textbf{40} (1932) 749 (1932).

\bibitem{orig2} W.\ Heisenberg
Z.\ Phys.\ {\bf 90} (1934) 209

\bibitem{orig3}
P.~A.~M.\ Dirac,
Math.\ Proc.\ Cam.\ Phil.\ Soc.\ {\bf 30} (1934) 150.



\bibitem{BialynickiBirula:1991tx}
  I.~Bialynicki-Birula, P.~Gornicki and J.~Rafelski,
  Phys.\ Rev.\  D {\bf 44} (1991) 1825.

\bibitem{Zhuang:1995pd}
  P.~Zhuang and U.~W.~Heinz,
  Annals Phys.\  {\bf 245} (1996) 311.

\bibitem{Hebenstreit:2010vz}
  F.~Hebenstreit, R.~Alkofer and H.~Gies,
  Phys.\ Rev.\  D {\bf 82} (2010) 105026.
  
\bibitem{Hebenstreit:2010cc}
  F.~Hebenstreit, A.~Ilderton, M.~Marklund and J.~Zamanian,
  Phys.\ Rev.\  D {\bf 83} (2011) 065007.

\bibitem{BialynickiBirula:2011uh}
  I.~Bialynicki-Birula and L.~Rudnicki,
  Phys.\ Rev.\  D {\bf 83} (2011) 065020.

\bibitem{Hebenstreit:2011wk}
  F.~Hebenstreit, R.~Alkofer and H.~Gies,
  arXiv:1106.6175 [hep-ph].

\bibitem{Blaschke:2011is}
  D.~B.~Blaschke, V.~V.~Dmitriev, G.~Ropke and S.~A.~Smolyansky,
  arXiv:1105.5397 [hep-ph].


\bibitem{Heinzl:2010vg}
  T.~Heinzl, A.~Ilderton and M.~Marklund,
  Phys.\ Lett.\  B {\bf 692} (2010) 250.

\bibitem{Honkanen:2010rc}
  H.~Honkanen, P.~Maris, J.~P.~Vary, S.~J.~Brodsky,
  Phys.\ Rev.\ Lett.\  {\bf 106 } (2011)  061603.

\bibitem{Meuren:2011hv}
  S.~Meuren and A.~Di Piazza,
  arXiv:1107.4531 [hep-ph].



\bibitem{Heinzl:2009nd}
  T.~Heinzl, D.~Seipt, B.~Kampfer,
  Phys.\ Rev.\  {\bf A81 } (2010)  022125.

\bibitem{Mackenroth:2010jk}
  F.~Mackenroth, A.~Di Piazza, C.~H.~Keitel,
  Phys.\ Rev.\ Lett.\  {\bf 105 } (2010)  063903.

\bibitem{Mackenroth:2010jr}
  F.~Mackenroth, A.~Di Piazza,
  Phys.\ Rev.\  {\bf A83 } (2011)  032106.


\bibitem{Tomaras:2000ag}
  T.~N.~Tomaras, N.~C.~Tsamis and R.~P.~Woodard,
  Phys.\ Rev.\  D {\bf 62} (2000) 125005.

\bibitem{Tomaras:2001vs}
  T.~N.~Tomaras, N.~C.~Tsamis and R.~P.~Woodard,
  JHEP {\bf 0111} (2001) 008.

\bibitem{Woodard:2001ai}
  R.~P.~Woodard,
  arXiv:hep-th/0110199.

\bibitem{Woodard:2001hi}
  R.~P.~Woodard,
  Nucl.\ Phys.\ Proc.\ Suppl.\  {\bf 108} (2002) 165.

\bibitem{Elze:1986qd}
  H.~T.~Elze, M.~Gyulassy and D.~Vasak,
  Nucl.\ Phys.\  B {\bf 276} (1986) 706.

\bibitem{Kibble:1975vz}
  T.~W.~B.~Kibble, A.~Salam, J.~A.~Strathdee,
  Nucl.\ Phys.\  {\bf B96 } (1975)  255-263.


\bibitem{Nikishov:1964zza}
  A.~I.~Nikishov and V.~I.~Ritus,
  Sov.\ Phys.\ JETP {\bf 19} (1964) 529.

\bibitem{Narozhnyi:1964}%
  N.~B.\ Narozhnyi, A.~Nikishov, and V.~Ritus,
  Zh.\ Eksp.\ Teor.\ Fiz.\ \textbf{47}, 930 (1964)
  [Sov.\ Phys.\ JETP \textbf{20}, 622 (1965)].

 
 
\bibitem{Narozhnyi:1976zs}
  N.~B.~Narozhnyi and A.~I.~Nikishov,
  Teor.\ Mat.\ Fiz.\  {\bf 26} (1976) 16.

\bibitem{Schwinger:1951nm}
  J.~S.~Schwinger,
  Phys.\ Rev.\  {\bf 82 } (1951)  664.
  
\bibitem{Soussa:2002ed}
  M.~E.~Soussa and R.~P.~Woodard,
  Phys.\ Rev.\  D {\bf 66} (2002) 085017.

\bibitem{Fried:2001ga}
  H.~M.~Fried, Y.~Gabellini, B.~H.~J.~McKellar and J.~Avan,
  Phys.\ Rev.\  D {\bf 63} (2001) 125001.

\bibitem{Avan:2002dn}
  J.~Avan, H.~M.~Fried and Y.~Gabellini,
  Phys.\ Rev.\  D {\bf 67} (2003) 016003.

\bibitem{Brezin:1970xf}
  E.~Brezin and C.~Itzykson,
  Phys.\ Rev.\  D {\bf 2} (1970) 1191.

\bibitem{Alkofer:2001ik}
  R.~Alkofer, M.~B.~Hecht, C.~D.~Roberts, S.~M.~Schmidt and D.~V.~Vinnik,
  Phys.\ Rev.\ Lett.\  {\bf 87} (2001) 193902.
  
\bibitem{Blaschke:2005hs}
  D.~B.~Blaschke, A.~V.~Prozorkevich, C.~D.~Roberts, S.~M.~Schmidt and S.~A.~Smolyansky,
  Phys.\ Rev.\ Lett.\  {\bf 96} (2006) 140402.

\bibitem{Narozhnyi:1970uv}
  N.~B.~Narozhnyi and A.~I.~Nikishov,
  Yad.\ Fiz.\  {\bf 11} (1970) 1072
  [Sov.\ J.\ Nucl.\ Phys.\  {\bf 11} (1970) 596]

\bibitem{Popov:1972}
  V.~S.~Popov,
  Sov.\ Phys.\ JETP {\bf 34} (1972) 709.

\bibitem{Popov:2001ak}
  V.~S.~Popov,
  JETP Lett.\  {\bf 74} (2001) 133 [Pisma Zh.\ Eksp.\ Teor.\ Fiz.\  {\bf 74} (2001) 151].

\bibitem{Hebenstreit:2009km}
  F.~Hebenstreit, R.~Alkofer, G.~V.~Dunne and H.~Gies,
  Phys.\ Rev.\ Lett.\  {\bf 102} (2009) 150404.

\bibitem{Dumlu:2010ua}
  C.~K.~Dumlu and G.~V.~Dunne,
  Phys.\ Rev.\ Lett.\  {\bf 104} (2010) 250402.

\bibitem{Volkov:1935}%
  D.~Volkov, Z.~Phys.\ \textbf{94}, 250 (1935).


\bibitem{Heinzl:2000ht}
  T.~Heinzl,
  Lect.\ Notes Phys.\  {\bf 572} (2001) 55.

\bibitem{Tajima}
  T.~Tajima and G.~Mourou, 
  Phys.\ Rev.\ ST Accel.\ Beams {\bf 5} (2002) 031301.
  
\bibitem{Hebenstreit:2011pm}
  F.~Hebenstreit,
  arXiv:1106.5965 [hep-ph].

\bibitem{Heinzl:2008rh}
  T.~Heinzl, A.~Ilderton,
  Opt.\ Commun.\  {\bf 282 } (2009)  1879-1883.
  
\bibitem{Heinzl:2003jy}
  T.~Heinzl,
  [hep-th/0310165].

\bibitem{BialynickiBirula:2011eg}
  I.~Bialynicki-Birula and L.~Rudnicki,
  arXiv:1108.2615 [hep-th].

\bibitem{Dunne:2002rq}
  G.~V.~Dunne,
  arXiv:hep-th/0207046.

\bibitem{Binosi:2003yf}
  D.~Binosi, L.~Theussl,
  Comput.\ Phys.\ Commun.\  {\bf 161 } (2004)  76-86.
  
\bibitem{Binosi:2008ig}
  D.~Binosi, J.~Collins, C.~Kaufhold, L.~Theussl,
  Comput.\ Phys.\ Commun.\  {\bf 180 } (2009)  1709-1715.

\bibitem{Fried:2001ur}
  H.~M.~Fried and R.~P.~Woodard,
  Phys.\ Lett.\  B {\bf 524} (2002) 233.

\end{thebibliography}
\end{document}